\documentclass{ws-p8-50x6-00}

\newcommand{\figsize}{18pc}

\begin{document}

\title{Thermodynamics of explosions}

\author{G. Neergaard}
\address{Niels Bohr Institute, Blegdamsvej 17, DK - 2100 Copenhagen, Denmark\\
and\\ 
Institute of Physics and Astronomy, University of Aarhus,
DK - 8000 Aarhus
\\E-mail: neergard@nbi.dk}

\author{J. P. Bondorf}
\address{Niels Bohr Institute, Blegdamsvej 17, DK - 2100 Copenhagen, Denmark\\
E-mail: bondorf@nbi.dk}

\author{I. N. Mishustin}
\address{Frankfurt University, D-60054, Germany\\
and\\
The Kurchatov Institute, Russian Research Center, 123182 Moscow, Russia\\
E-mail: mishustin@nbi.dk}

\maketitle

\abstracts{
We present our first attempts to formulate a thermodynamics-like
description of explosions.
The motivation is partly a fundamental interest in non-equilibrium 
statistical physics, partly the resemblance of an explosion
to the late stages of a heavy-ion collision.
We perform numerical simulations on a microscopic model of 
interacting billiard-ball like particles,
and we analyse the results of such simulations trying to
identify collective variables describing the 
degree of equilibrium during the explosion.}

\section{Introduction}
The assumption of thermodynamic equilibrium at an intermediate stage 
of a heavy-ion collision is often incorporated in models
of the colliding nuclear matter.
These models range from statistical models of nuclear multifragmentation 
to the fluid dynamical models of the quark gluon plasma. 
In contrast, microscopic models of molecular dynamics type 
(e.g.\ RQMD, FMD and NMD), which are based upon constituent interactions, 
do not contain this assumption.
Such models are appropriate for testing to what extent 
thermodynamic equilibrium is actually achieved.
And if it is not, the application of thermostatic concepts 
such as temperature and entropy becomes questionable. 
In this study we employ a very simple model,
and focus on the thermodynamic or ``overall'' description
of the system.

\section{The model}
Our model consists of a number $A$ of identical balls of radius $r_{hc}$
having mass $m$. 
They perform classical non-relativistic hard-sphere 
scatterings, conserving energy, momentum and angular momentum.
Initially the $A$ balls are placed randomly within a sphere 
of radius $R=R_0 A^{1/3}$, 
and the initial velocities are chosen as a
superposition of thermal (Maxwell-Boltzmann) and collective motion.
We use a spherically symmetric Hubble-like flow field for the 
initial collective motion:
\be
\vec{v}(\vec{r}) = -v_{0f}\,\vec{r}/R
\label{hubbleflow}
\ee
where $v_{0f}$ is a model parameter, $v_{0f}>0$ for ingoing flow
and $v_{0f}<0$ for outgoing flow.
We fix the total energy $E=E_{fl}+E_{th}$,
and vary the fraction $\eta$ of the flow energy, $\eta=E_{fl}/E$,
where $E_{fl}=\frac{m v_{0f}^2}{2 R^2}\sum_{i=1}^A\vec{r}_i^2$ 
and $E_{th}$ are the flow energy and the thermal energy,
respectively.
Because of the way in which the system is built up, these energies
will fluctuate from event to event with a relative uncertainty 
of the order of $A^{-\frac{1}{2}}$.   

In our simulations we have chosen nuclear-scale parameters:
$m=940$~MeV, $r_{hc}=0.5$~fm, $R_0=1.2$~fm, $0 \leq v_{0f} \leq 0.5$ 
(in units of the velocity of light, $c=1$),
but since the behavior of the model only depends on 
the two combined parameters
$mv_{0f}^2$ and $r_{hc}/R$,
the choice of nuclear scale is not crucial.
We choose $A=50$, so with these parameters 
the initial radius of the system is $4.2$~fm.

We focus on four different types of event:
$\bullet$~'th20':~The particles are started in $100\%$ thermal motion 
inside a spherical container of radius $4.2$~fm, 
at $t=20$~fm/c the container walls are removed.
$\bullet$~'in':~$100\%$ ingoing flow. 
After interacting, the particles will move out again.
This implosion-explosion process is intended to simulate some features of
a heavy-ion collision.
$\bullet$~'50/50':~$50\%$ thermal motion + $50\%$ outgoing flow, 
simulating an explosion from a non-thermalized state.
$\bullet$~'100out7':~100\% outgoing flow inside a spherical container 
of radius $7$~fm.

The results are averaged over an ensemble of $20$ events of
each kind.

\section{Thermodynamic considerations}
It is clear that we cannot use ordinary thermodynamics 
(or its well-known extensions to small systems~\cite{hill} 
or to small deviations
from equilibrium~\cite{zubarev}) for the 
description of the overall behavior of our model.
First, it is not clear that equilibrium prevails, even locally.
Indeed we wish to investigate to what extent equilibrium is reached
in the course of an implosion-explosion process.
Second, our system has no fixed volume, it expands freely into the vacuum.
It is the combination of these two facts, no temperature and no volume,
that makes our approach different from much previous work 
on the subject~\cite{prevstud}.

Equilibrium thermodynamics is linked to the 
motion of the individual constituents making up the macroscopic
system via the entropy~\cite{kennard,ll}.
A natural starting point for the investigation of the overall,
{\it i.e.}\ the ``thermodynamic'', behavior of our system
is therefore to apply an expression similar to the entropy, 
but in a way that makes sense in this highly non-equilibrium system.

To study one-body observables, we
reduce the $6A$ dimensional phase-space of the $A$ particles 
to $6$ dimensions in the standard way~\cite{kennard}.
Then we introduce a finite grid in the reduced phase-space,
dividing each of the $6$ axes into $D$ segments.
Instead of working with a fixed grid in phase-space, which would 
give us the usual entropy\footnote{in the limit $D\rightarrow \infty$
in an ensemble of infinitely many events}, 
we let the entire grid expand or contract along with the swarm of points
in phase-space in a uniform way:
The outer grid edge follows the outermost point,
the boxes are of equal size, and
the number of boxes is kept fixed,
thus the physical size ({\it e.g.}\ in units of $\hbar^3$)
of each box in phase-space varies with time.
This is to deal with the {\em no volume} problem, we mentioned above.
We then introduce the {\it pseudo-entropy} as
\begin{equation}
\Sigma = -\frac{1}{\xi} \sum_i p_i \log(p_i)
\label{pS}
\end{equation}
where
\begin{equation}
p_i=\frac{{number\ of\ points\ in\ box}\ i}{total\ number\ of\ points\ in\ phase\ space}
\label{plogp}
\end{equation}
and $\xi$ is a normalization constant.
We choose $\xi$ as the theoretical maximum value of $-\sum_i p_i \log(p_i)$
so that $\Sigma \in [0,1].$
In the current set-up, for the case of $N$ points in a reduced phase-space 
divided into $D^6$ boxes, $\xi$ is the smaller number of 
$\log(N)$ and $6\log(D)$.
We refer to the quantity $\Sigma$ as the pseudo-entropy
instead of entropy, since some important features of the entropy, 
{\it e.g.}\ that it increases, are not retained in this formulation.
Nevertheless, we shall see that $\Sigma$ has some nice properties, 
including that of characterizing the degree of equilibrium.
For a system of fixed volume in equilibrium, $\Sigma$ is the usual
one-body entropy (apart from normalization).

\section{Results}
In the calculations presented here, 
we have chosen $D=7$, so the $6$ dimensional
phase-space is divided into $7^6$ boxes. 
Because we are dealing with a small number of events (typically $20$) 
in the ensembles, 
the precise values of $\Sigma$ depend on the choice of grid.
We have, however, verified that $\Sigma$ behaves qualitatively
similar to what is shown here over a range of grid-sizes,
varying $D$ between $2$ and $9$.

\begin{figure}[h]
\epsfxsize=\figsize
\epsfbox{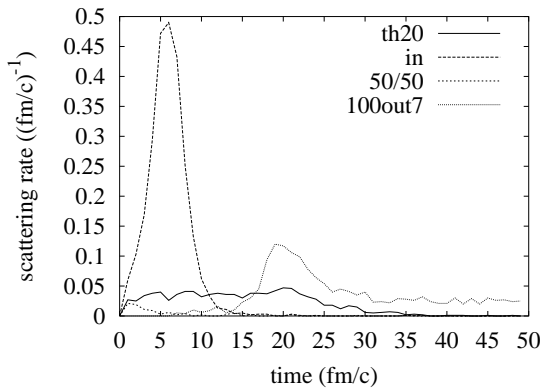}
\caption[]{The scattering rate (number of scatterings per particle
per time unit) for the four cases mentioned in the text.
In the implosion-explosion event ('in')
practically all particles scatter around the time $t \simeq 6$~fm/c.
This is the time when the system is maximally compressed. 
Then, as the expansion begins, the scattering rate decreases 
until $t \simeq 15$~fm/c, when interactions have essentially ceased.
In the '100out7' case the particles start to hit the container wall 
at $t\simeq 5$~fm/c (the scatterings against the container walls 
are not counted here), and the peak in the scattering rate at 
$t\simeq 20$~fm/c results from particles moving back after hitting
the wall and scattering against other particles still on their way out.
\vspace{-4 mm}
\label{fig:scatrate}}
\end{figure}

To give an idea of the dynamics and the timescales, we show in 
Fig.~\ref{fig:scatrate} the scattering rate.

\begin{figure}[h]
\epsfxsize=\figsize
\epsfbox{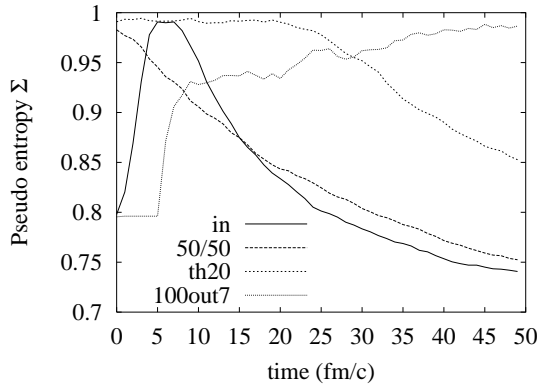}
\caption[]{The pseudo-entropy for the four cases described in the text.
This variable seems to quantify the degree of equilibrium in the system, 
$\Sigma=1$ characterising an equilibrium state.
\vspace{-6mm}
\label{fig:pseudos}}
\end{figure}

Fig.~\ref{fig:pseudos} shows how the pseudo-entropy behaves
in each of the four cases.
In the 'th20' case, the pseudo-entropy $\Sigma \simeq 1$
as long as the particles are in equilibrium at fixed volume 
inside the container. Then at $t=20$~fm/c, when the container is
removed and the system starts to expand, the pseudo-entropy
decreases, reflecting the fact that the system goes 
out of equilibrium\footnote{
The particles stay almost {\it thermalized}, though, in the sense that 
they retain their Maxwell-Boltzmann velocity distribution.
But they are certainly not in {\it equilibrium}, since this means that 
the phase-space distribution is independent of time.}.

In the case '100out7', where the particles are started 
in an extreme non-equilibrium situation, the pseudo-entropy
is low ($\Sigma=0.8$ is a low value in this context), 
but increases towards $\Sigma=1$ as the scatterings equilibrize
the system.  
By comparison with Fig.~\ref{fig:scatrate} one can see
that the first jump in $\Sigma$ at $t\simeq 5$~fm/c 
is due to particles scattering against the container wall
(when particles hit the wall their velocity is reversed,
so in this process many new states in phase-space are being populated),
and the second ``jump'' around $t\simeq 20$~fm/c is due to
the many particle-particle scatterings around this time.

The interesting case is the implosion-explosion ('in') scenario,
since here we do not know in advance if the system 
reaches a state of equilibrium or not.
From the fact that the pseudo-entropy in Fig.~\ref{fig:pseudos}
reaches a value $\Sigma \simeq 1$, the same value as the known
equilibrium case 'th20',
we infer that the system {\it is} in a state of equilibrium
around $t \simeq 6$~fm/c (which is also the time of
maximum compression).
We have checked that the speed
distribution of the particles becomes nearly 
Maxwellian from $t \simeq 6$~fm/c 
with a temperature in the compressed state of $47$~MeV, 
which is also the theoretical value of the temperature 
assuming that all of the initial flow energy is converted to 
thermal energy.

\begin{figure}[h]
\epsfxsize=\figsize
\epsfbox{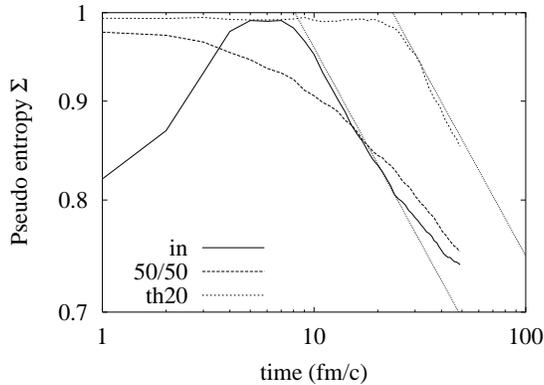}
\caption[]{The pseudo-entropy in a log-log plot, together with the 
functions: $1.52\,t^{-0.2}$ and $1.87\,t^{-0.2}$.
The decrease of $\Sigma$ during the initial expansion of the system
seems to follow a power law when a state of equilibrium {\it was} present,
in contrast to the case '50/50' 
(intended to simulate the expansion from a not-fully thermalized state)
which does not show this behavior.
At later times $\Sigma$ decreases less rapidly and turns over 
to approach a finite limiting value,
one sees the beginning of this behavior at the curve 'in'.
\vspace{-6mm}
\label{fig:logpseudos}}
\end{figure}

Another interesting feature of the pseudo-entropy is that it seems
to decay in a characteristic fashion when the system expands from
a state of equilibrium, see Fig.~\ref{fig:logpseudos}.

\section{Conclusions}
In this note we address the problem of thermodynamic equilibration 
in the context of heavy-ion collisions.
We have defined a variable inspired by the entropy
which, at least for the cases we have considered, 
seems to characterize the degree of equilibrium in an {\it a priori} 
highly non-equilibrium process such as an explosion.
Now, more theoretical work needs to be done in order to understand
{\it why} $\Sigma$ behaves in this seemingly interesting way.

\section*{Acknowledgements}
\vspace{-4mm} 
This work was in part supported by the Danish Natural Science Research Council.
GN thanks the Niels Bohr Institute for kind hospitality
and The Leon Rosenfeld Foundation for financial support during
the preparation of this work.
JPB thanks the Nuclear Theory Group at GSI, where part of this work was done.
INM thanks The Humbold Foundation for financial support.

\end{document}